\documentstyle[sprocl]{article}

\def\be{\begin{equation}}
\def\ee{\end{equation}}
\def\bea{\begin{eqnarray}}
\def\eea{\end{eqnarray}}

\begin{document}

\title{BRICK WALL AND QUANTUM STATISTICAL ENTROPY OF BLACK HOLE}

\author{S.N. SOLODUKHIN}

\address{Dept. of Physics, University of Waterloo, 
Waterloo,\\ ONT N2L 3G1, CANADA}

\maketitle\abstracts{We discuss the statistical-mechanical entropy of black hole
calculated according to 't Hooft. It is argued that in presence of horizon
 the statistical mechanics of 
quantum fields depends on their UV behavior. The ``brick wall''  model was shown
to provide a correct description when the ``brick wall'' parameter is  less
 than any UV cut-off.
}

Since Bekenstein introduced the thermodynamical analogy in black hole physics
 and Hawking discovered  thermal radiation from a black hole confirming 
this analogy \cite{1}, it is an intriguing problem as to 
what degrees of freedom are 
counted by the entropy of a black hole.
Equivalently, what (if any) statistical mechanics is responsible for the Bekenstein-Hawking 
entropy? 

According to 't Hooft \cite{4} the statistical-mechanical entropy $S_{SM}$ arises
 from a thermal bath of quantum fields propagating outside the horizon.
It should be noted that every calculation of statistical entropy encounters  
the problem of dealing with the very peculiar behavior of the 
physical quantities near the
horizon where they typically diverge. To remove these divergences 't Hooft 
introduced ``brick wall'': a fixed  boundary near the horizon within which 
the quantum field does not propagate. Essentially,  this procedure (as it 
formulated in \cite{4}) must be implemented in addition
to the removing of standard ultra-violet divergences. 
Contrary to statistical entropy the thermodynamical entropy of black hole is finite
(after UV-renormalization)  quantity not possessing any kind of
the ``brick wall'' divergence. This fact has inspired to argue in \cite{13}
that black holes provide us with a unique exapmle of a specific system 
for which these entropies
do not necessarily coincide. 

However, dealing with {\it quantum} field, that is system having infinite number
of degrees of freedom, we have to take into account that not all of these
freedoms are, in fact, physical. Indeed, the high energy modes lead to 
unphysical infinities and must be subtracted. The well-defined UV-renormalization
is invariant procedure of such a subtraction. Therefore, formulating 
the statistical mechanics for a quantim field we need in the similar
subtraction of unphysical modes which must be excluded in the 
statistical ensemble.
How this really happens was demonstrated in \cite{14} by using 
the Pauli-Villars (PV) regularization scheme. It consists in introducing 
a number of fictitious fields (regulators)
of different statistics and with very large masses. Remarkably, this procedure 
not only yields the standard UV regularization but automatically implements a 
cut-off for the entropy calculation
allowing to remove ``brick wall''.

The interplay of ``brick wall'' and UV regularization is easily seen for two-dimensional massless scalar field.
Applying Pauli-Villars regularization in two dimensions
one needs to introduce a set of fictitious fields with very large masses:
two anticommuting scalar fields with  mass $\mu_{1,2}=\mu$
and one commuting field with   mass $\mu_3=\sqrt{2}\mu$.
Consider the free energy of the  ensemble of the original scalar field and
 regulators with an inverse temperature $\beta$:
\begin{equation}
\beta F= \sum_n \ln (1-e^{-\beta E_n})
\label{C}
\end{equation}
Note that energy $E_n$ in (\ref{C}) is defined with respect to Killing vector
 $\partial_t~~
(\tau=\imath t )$ and fields are expanded  as $\phi=e^{\imath E t} f(x)$.
 Therefore, $\beta$ in (\ref{C}) is related with temperature $T$ measured at $x=L$
as $T^{-1}=\beta g^{1/2}(L)$. The relevant density matrix is $\rho=\sum_n \phi_n
\phi^*_n e^{-\beta E_n}$, where $\{ \phi_n \}$ is basis of eigen-vectors.
One should take into account that for the regulator fields the Hilbert space hase indefinite
metric and hence a part of regulators contributes with minus sign.

The free energy (\ref{C}) can be determined for the arbitrary
black hole metric $ds^2=-g dt^2+g^{-1}dx^2$ without reference to the precise form of the metric
 function $g(x)$. 
Repeating the calculation of ref.\cite{14} in this 2D case and applying WKB 
approximation  we finally get 
\begin{eqnarray}
&&F=-{1\over \pi} \int^\infty_0{dE \over e^{\beta E}-1} \int^L_{x_++h}{dx \over g(x)}
( E-2(E^2-\mu^2 g(x))^{1/2} \nonumber \\
&&+(E^2-2\mu^2 g(x))^{1/2} )~~.
\label{7}
\end{eqnarray}
It should be noted that the WKB approximation for the original massless scalar
 field is really exact.
We introduced in (\ref{7}) a  ``brick wall'' cut-off $h$. In fact, one can
 see that divergences at small $h$
are precisely cancelled in (\ref{7}) between the original scalar and the
 regulator fields.
This is 2D analog of the mechanism discovered in \cite{14}.
So one can remove the cut-off in (\ref{7}). However we will keep it  
arbitrarily small in the process of calculation of
 separate terms entering in (\ref{7}).

It is straightforward to compute the contribution of the original massless field in
 (\ref{7}). For computation of the regulator's contribution take  the 
fixed $E$ and consider the integral:
\begin{equation}
I[\mu ]= \int^{L_E}_{x_++h}{dx \over g(x)}(E^2-\mu^2 g(x))^{1/2}~~,
\label{8} 
\end{equation}
where integration is doing from the horizon ($x_++h$) to distance $L_E$ defined from equation
$g(L_E)={E^2 \over \mu^2}$.  It is clear that when $\mu$  grows
$L_E$ becomes closer and closer to ($x_++h$). So, considering 
 limit of large $\mu$ we conclude that integral (\ref{8}) is concentrated
near the horizon where we have: $g(x)={4\pi \over \beta_H}(x-x_+)=({2\pi \rho \over
\beta_H})^2,~~{dx\over g}={\beta_H \over 2\pi}{d\rho \over \rho}$ and the new
radial variable $\rho$ now runs from  $\epsilon=\sqrt{\beta_H h \over \pi}$ to $ ({E\beta_H \over 2\pi \mu})$.
The integral (\ref{8}) then in the limit of small $\epsilon$  reads:
\begin{eqnarray}
&&I[\mu ]=\mu \int_\epsilon^{E\beta_H \over 2\pi \mu}{d \rho \over \rho} \sqrt{({E\beta_H \over 2\pi \mu})^2-\rho^2} \nonumber \\
&&=-{(E\beta_H) \over 2\pi}(1+{1\over 2} \ln 2 +\ln ({\mu\epsilon\pi \over E \beta_H}))~~.
\label{10}
\end{eqnarray}
This is the key identity allowing computation of the free energy (\ref{7}).
Omitting details which are rather simple  the result is
\begin{equation}
F=-{1\over 12}[ {\beta_H \over 2 \beta^2} \int^L_{x_+}{dx \over g}
({4\pi \over \beta_H} -g')+{\beta_H \over \beta^2} \ln ( \mu \beta g^{1/2}(L))]
+{\beta_H \over \beta^2} C~~,
\label{11}
\end{equation}
where we removed the brick wall cut-off and used that $
\int^\infty_0{dx x \over e^x-1}={\pi^2 \over 6}~~.$  The statistical-mechanical free energy (\ref{11}) is really an off-shell quantity (see \cite{14}) defined for  
arbitrary black hole metric  and $\beta$ not necessarily equal to $\beta_H$.

Calculating now entropy $S_{ST}=\beta^2\partial_\beta F$ and putting $\beta=\beta_H$
we obtain:
\begin{equation}
S_{ST}={1\over 12} \int^L_{x_+}{dx \over g}({4\pi\over \beta_H}-g')+ {1\over 6}\ln (\mu \beta_H g^{1/2}(L)) +C~~,
\label{12}
\end{equation}
where $C$ is some numerical constant not depending on $\mu$ or metric $g(x)$.
As was demonstrated in \cite{S}  $S_{ST}$(\ref{12})  exactly coincides with thermodynamical
entropy of quantum field.
So, at least this part of the thermodynamical entropy has the 
statistical meaning. 

It is important to note the crucial interplay 
of two different 
limits $h\rightarrow 0$ (brick wall)
and $\mu^{-1}\rightarrow 0$ (UV-regulatorization). If one takes the limit
$\mu^{-1}\rightarrow$ first one obtains that contribution of the regulators in the
free energy (\ref{7}) completely vanishes. On then gets the quantities which are functions of the brick wall parameter $h$ and divergent in the limit $h\rightarrow 0$.
These are that quantities calculated in \cite{13}. Elimination of their divergence (with respect to limit $h\rightarrow 0$) might require some subtraction procedure proposed in 
\cite{13}. Note, that in this regime the "brick wall" is treated as real boundary staying at
{\it macroscopical} distance $h$ from the horizon with $h$ being larger than any
UV cut-off $\mu^{-1}$. However, in this case, this is no more a black hole.

The situation is different if we consider "brick wall" as an fictitious imaginary boundary
with $h$ being smaller any scale $\mu^{-1}$ of UV cut-off. Then the "brick wall"
divergences are eliminated by the standard UV-regularization and the
UV-regulators do contribute to the free energy and entropy. This contribution is concentrated at the horizon. It leads to appearance of
additional terms ($\int g^{-1} g'$) in the entropy (\ref{12}) that are finite after renormalization. It is worth noting that mechanism of this phenomenon  is similar to that of the conformal anomaly.
This similarity is not occasional since the result for the statistical entropy
(\ref{12}) occurs to coincide with the thermodynamical expression 
which is indeed 
originated from the conformal anomaly of the Polyakov-Liouville action.
We do not have this phenomenon in the statistical mechanics on space-time without
horizons where the statistical entropy was proved to be conformal invariant and not
dependent on UV cut-off (see \cite{D}). This is easily seen from our analysis.
Indeed, in this case we have $g(x) \geq g_0>0$ everywhere and for large
UV cut-off $\mu> \mu_0={E\over g^{1/2}_0}$ contribution of the regulators
disappears in the free energy (\ref{7}).
Thus, in the presence of horizons the statistical mechanics of quantum fields depends
on their UV behavior. The UV-regulators lead to non-trivial contribution to
statistical entropy that is finite after renormalization. Unfortunately,
the straightforward generalization of this result on higher dimensions
meets the still open problem of statistical description of the non-minimally
coupled conformal matter.

This work was supported by NATO and the Natural Sciences and Engineering Research
Council of Canada.

\end{document}